\documentstyle [aps,preprint,epsf]{revtex}
\def\bx{{\mbox{\boldmath $x$}}}

\def\b0{{\mbox{\boldmath $0$}}}
\def\bPsi{{\mbox{\boldmath $\Psi$}}}
\def\sb{{\sf b}}
\def\sd{{\sf d}}
\begin{document}
\title{Coherent effects in magneto-transport through Zeeman-split
levels}
\author{S. A. Gurvitz$^{1,2}$ {\rm ,} D. Mozyrsky$^2$ {\rm ,}
and G. P. Berman$^2$\\
$^1$Department of Particle Physics, Weizmann Institute of Science,
Rehovot 76100, Israel\\ $^2$Theoretical Division and CNLS, Los
Alamos National Laboratory, Los Alamos, NM 87545, USA\\}
\date{\today}
\maketitle
\begin{abstract}
We study non-equilibrium electronic transport through a quantum
dot or an impurity weakly coupled to ferromagnetic leads. Based on
the rate equation formalism we derive the noise spectra for the
transport current. We show that, due to quantum interference
between different spin components of the current, the spectrum
develops peaks or dips at frequencies corresponding to the Zeeman
splitting in the quantum dot. A detailed analysis of the spectral
structure of the current is carried out for noninteracting
electrons as well as for the regime of Coulomb blockade.
\end{abstract}
\newpage

\section{Introduction}

Resonant transport through a quantum dot (or impurity) has been
investigated in numerous publications. Yet no special attention
has been paid to quantum interference effects in this process.
These effects can generate oscillations in the resonant current
through two (or more) levels of an impurity. These oscillations
are similar to the well-known quantum interference effects in the
two-slit experiment, and, in turn, produce a peak or a dip in the
current power spectrum, depending on the relative phase of the two
levels carrying the current\cite{ieee}. This feature can be
experimentally observed in time-resolved measurements of transport
currents \cite{man,dur}. It has been previously argued that the
quantum interference effect can explain modulation in the
tunneling current at the Larmor frequency in scanning tunneling
microscope (STM) experiments\cite{moz}.

In this paper we investigate the interference effects in polarized
magneto-transport. Conductance and I-V curves for spin dependent
transport through quantum dots has recently been studied in
several publications\cite{mar,br,ped,gor,wey}. Here we study the
time dependent properties of transport currents. In particular we
study the effects related to interference between different spin
components of the currents.

These effects can be described schematically as follows. Consider
the polarized resonant current from the left reservoir (emitter)
to the right reservoir (collector) through a single level of a
quantum dot (impurity) in the presence of an external magnetic
field. This field would split the resonant level of the dot into a
Zeeman doublet, Fig.~1. Let us assume that the polarization axis
of electrons in the emitter ({\bf n}) is different from that of
the external magnetic field ({\bf {\={n}}}). Then a spin-polarized
electron from the emitter enters into a superposition of the
``spin-up" and the ``spin-down" states of the Zeeman doublet,
Fig.~1. As a result, the electron wave function in the collector
has two components corresponding to different energies of the
doublet. Yet these components are orthogonal since they correspond
to different spin components and therefore cannot interfere. If,
however, there is an additional spin-flip process in the
transition between the dot and the collector, the two spin
components interfere in the collector current. Again, this takes
place if the polarization in the collector ({\bf n$^{\prime}$}) is
different from that in the quantum dot ({\bf{\={n}}}), Fig.~1.
Thus the system operates as a two -path interferometer, where the
phase difference between the two paths, i.e., through the upper
and the lower spin states in the dot, contributes to the dynamical
(or spectral) properties of the collector current.

\vskip1cm
\begin{minipage}{13cm}
\begin{center}
\leavevmode \epsfxsize=8cm \epsffile{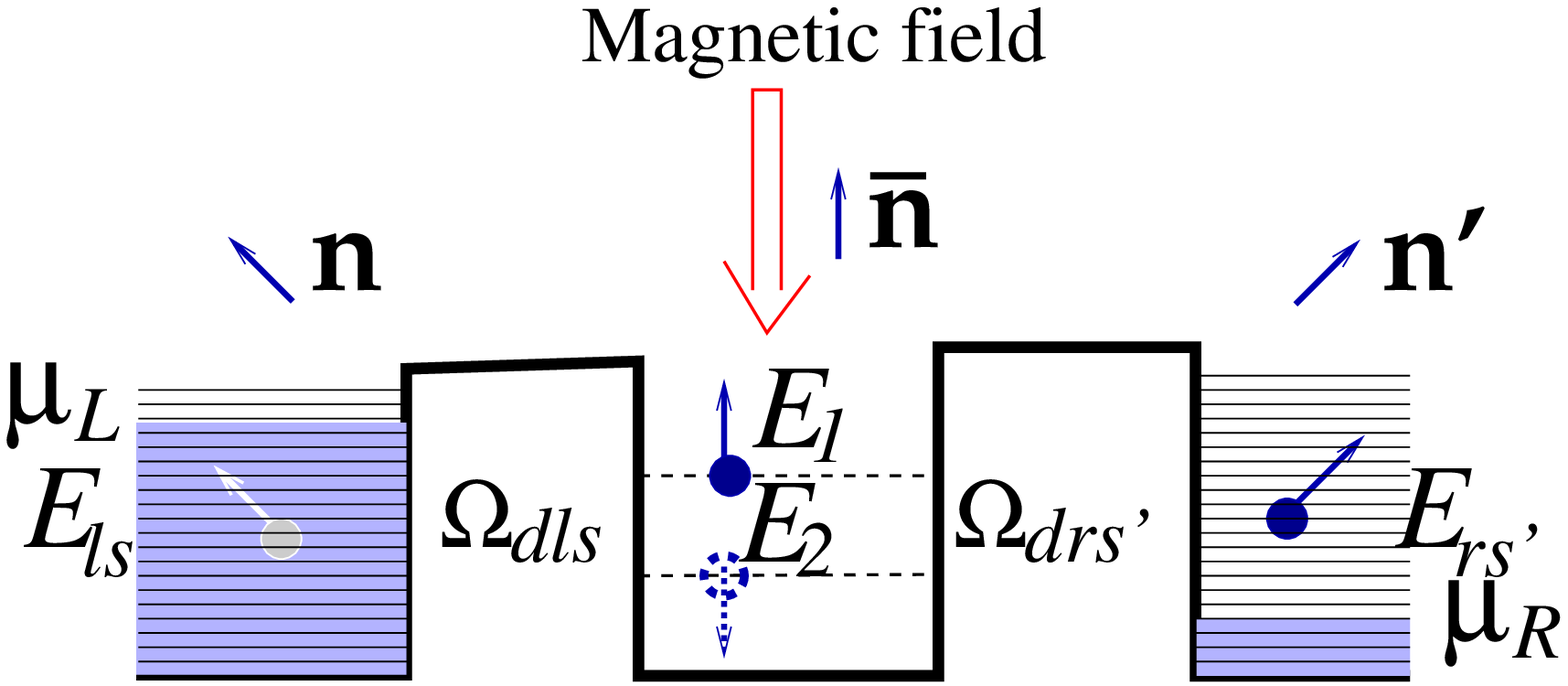}
\end{center}
{\begin{small} Fig.~1. Resonant tunneling of a polarized electron
through a quantum dot. Here the $\Omega$'s denote the tunneling
transition amplitudes between the reservoir states and the Zeeman
doublets ($E_{1,2}$) of the quantum dot. $\mu_{L,R}$ are the
chemical potentials in the left and right reservoirs. The unit
vectors {\bf n}, {\bf \={n}} and {\bf n$^{\prime}$} show the
polarization axes in the emitter, quantum dot, and the collector,
respectively.
\end{small}}
\end{minipage} \\

These interference effects can be realized experimentally in a
heterostructure with a quantum dot sandwiched between the two
ferromagnetic leads with easy axes different from those in the
dot. A similar set up can be implemented in other systems, such as
self-assembled quantum dots\cite{aw}, ultrasmall
particles\cite{des}, carbon nanotubes\cite{nyg} and single
molecules\cite{pasu}. These systems are likely to find prominent
technological applications, including random excess memory and
magnetic sensors due to giant magnetoresistance effect\cite{an}.
The value of Zeeman splitting in the dot is controlled by the
external magnetic field. The leads are assumed to be thin magnetic
films, so that the magnetic field inside the leads is pinned
parallel to the films and therefore the magnetization in the leads
remains unaffected by the application of a relatively small
external magnetic field.

The plan of this paper is as follows. In Sect. II we introduce the
model and describe in general the rate equation approach for
calculations of the resonant current and its noise spectrum. This
approach has been obtained directly from the many-body
Schr\"odinger equation describing the entire many-particle
system\cite{g,gg} and allows one to treat the magneto-transport
through quantum dot systems in the most simple and precise way. In
Sect. III we consider the case non-interacting (of weakly
interacting) electrons. We explain in detail how the rate
equations are used to calculate the polarized current and the
noise spectrum both for ferromagnetic and non ferromagnetic leads.
The results for the time-dependent polarized current are compared
with the results of the single-particle model, presented in the
Appendix. We explicitly demonstrate how the polarized current
exhibits oscillations due to interference effects. In Sect. IV we
concentrate on the interacting case. In particular we derive the
current spectra in the presence of a Coulomb blockade in the dot.
We consider separately the collector and the emitter currents, as
well as the circuit current. In Sect. V we summarize our
calculations and briefly discuss the potential implications of our
results on the noise spectroscopy of quantum dots.

\section{Many-body description.}

Consider the polarized transport of non-interacting electrons
through a quantum dot in the external magnetic field, Fig.~1. The
polarization axis of an electron inside the dot ({\bf \={n}}) is
different from those in the right and left reservoirs ({\bf n} and
{\bf n$^{\prime}$}). The tunneling Hamiltonian describing this
system can be written as
\begin{eqnarray}
&&H=\sum_{l,s}E_{ls}a_{ls}^\dagger a_{ls}
+\sum_{d=1,2}E_{d}a_{d}^\dagger a_{d}
+\sum_{r,s'}E_{rs'}a_{rs'}^\dagger a_{rs'}
\nonumber\\
&&~~~~~~~~~~~~~~~~~~~~~~~~ +\left
(\sum_{d,l,s}\Omega_{dls}a_{ls}^\dagger a_{d}
+\sum_{d,r,s'}\Omega_{drs'}a_{rs'}^\dagger a_{d}+H.c\right
)+U_Ca^\dagger_{1}a_{1}a^\dagger_{2}a_{2} \, , \label{a1}
\end{eqnarray}
where the spin indices, $s,s'=\pm 1/2$ are related to different
quantization axes ({\bf n} and {\bf n'}, Fig. 1), and
$E_{ls},E_{rs'}$ denote the energy levels in the reservoirs. The
Zeeman splitting of the dot is denoted by $E_d$, where $d=1,2$.
The last two terms describe the tunneling transitions between the
reservoirs and the dot states generated by both the tunneling
couplings $\Omega$ and the Coulomb repulsion of two electrons
inside the dot.

All parameters of the tunneling Hamiltonian (\ref{a1})
are related to the initial microscopic description of the system
in the configuration space ($\bx $).
For instance, the coupling $\Omega_{dls}$
is given by the Bardeen formula\cite{bar}
\begin{equation}
\Omega_{dls}=-{1\over 2m}\int_{\bx\in\Sigma_{l}}\phi_{d}(\bx )
\stackrel{\leftrightarrow}\nabla\chi_{ls}(\bx )d\sigma\ ,
\label{aa3}
\end{equation}
where $\phi_{d}(\bx )$ and $\chi_{ls}(\bx )$ are the electron wave
functions inside the dot and the reservoir, respectively and
$\Sigma_l$ is a surface inside the potential barrier that
separates the dot from the left reservoir. Since the spin
quantization axes in the dot and in the leads differ from each
other, the transition matrix elements ($\Omega$'s) in
Eq.~(\ref{aa3}) depend on the relative angles between the dot and
the lead polarization axes ($\theta_L$ and $\theta_R$ for left and
right leads respectively). The simplest form of the couplings that
respects SU(2) symmetry is
\begin{equation}
  \Omega_{dls}=\Omega_l \sd^{(1/2)}_{s,s_d}(\theta_L)~~~{\mbox{and}}~~~
\Omega_{drs'}=\Omega_r \sd^{(1/2)}_{s_d,s'}(\theta_R)\, ,
\label{a33}
\end{equation}
\noindent where $s_d=\pm 1/2$ denotes the electron spin inside the
dot, Fig.~1, $\sd^{(1/2)}(\theta)$ is spin rotation matrix,
\begin{equation}
\sd^{1/2}(\theta ) =
\pmatrix{\cos{\theta\over2}&\sin{\theta\over2}\cr
  -\sin{\theta\over2}&\cos{\theta\over2} }\, ,
   \label{a34}
\end{equation}
and $\Omega_{l/r}$ is spin-independent part of the couplings.
Neglecting the energy dependence of these couplings,
$\Omega_{l,r}=\Omega_{L,R}$, one can relate them to the partial
widths (tunneling rates) as
$\Gamma_{L,R}=2\pi\Omega_{L,R}^2\rho_{L,R}$, where $\rho_{L,R}$ is
density of the states in the left (right) reservoir.

In the case of large bias, $|E_{1,2}-\mu_{L,R}|\ll \Gamma_{L,R}$,
the many-body Coulomb repulsion effects in the magneto-transport
can be accounted for in the most simple and precise way by using
the modified Bloch-type equations for the reduced density
matrix\cite{g,gg}. These equations can be derived from the
many-body Schr\"odinger equation by integrating out the reservoir
states in the limit of weak or strong Coulomb repulsion,
$U_C\ll\mu_L-E_1$ or $U_C\gg\mu_L-E_1$, without any stochastic or
other approximations. In addition these equations are very useful
for evaluating of the shot-noise power spectrum.

In order to apply our method we first redefine the vacuum state,
$|\b0\rangle$, by identifying it with the initial state of the
entire system. For instance we can identify it with empty dot,
with the emitter and collector filled up to the chemical
potentials $\mu_{L,R}$, respectively. We also assume that the
electrons in the emitter are polarized along the {\bf
n}-direction, Fig.~1. The many-body wave function, describing the
entire system can be written in the most general way as
\begin{eqnarray}
&&|\bPsi (t)\rangle =\left
[b_0(t)+\sum_{d,l,s}b_{dls}(t)a^\dagger_{d}a_{ls}
+\sum_{l,s,r,s'}b_{rs'ls}(t)a^\dagger_{rs'}a_{ls}\right.
\nonumber\\
&&~~~~~~~~~~~~~~~~~~~~~~~
\left. +\sum_{l,s,\bar l,\bar s} b_{12ls\bar l\bar s}(t)
  a^\dagger_1a^\dagger_2a_{ls}a_{\bar l\bar s}
  +\sum_{d,l,s,\bar l,\bar s,r,s'}
  b_{drs'ls\bar l\bar
    sr}(t)a^\dagger_{d}a^\dagger_{rs'}a_{ls}a_{\bar l\bar s}
  +\cdots\right ]
|\b0\rangle \, ,
\label{b1}
\end{eqnarray}
where $d=\{1,2\}$ denotes a state with one electron in the dot and
$ls(rs')$ denote the electron level in the emitter (collector).
The amplitudes $b_\alpha (t)$ of finding the entire system in the
state ``$\alpha$'' are obtained from the Schr\"odinger equation,
$i\partial_t|\Psi (t)\rangle ={\cal H}|\Psi (t)\rangle$ with the
initial condition $b_\alpha (0)=\delta_{\alpha,0}$

Let us introduce the (reduced) density matrix
$\sigma_{jj'}^{m,n}(t)$, where $n,m$ denote the number of electrons
arriving the right reservoir with the spin components $s'=\pm 1/2$,
respectively, Figs.~1,2. The lower indices, $j,j'$ denote the
discrete states of the quantum dot. For instance, in the case of
non-interacting (or weakly interacting) electrons
$j,j'=\{0,1,2,3\}$,  Fig.~2. This density matrix,
$\sigma_{jj'}^{m,n}(t)$ can be easily constructed from the
amplitudes $b(t)$, Eq.~(\ref{b1}). For example,
\begin{eqnarray}
&&\sigma_{00}^{0,0}(t)=|b_0(t)|^2,~~~\sigma_{11}^{0,0}(t)=\sum_{l}|b_{1l1/2}(t)|^2,
~~~\sigma_{22}^{0,0}(t)=\sum_{l}|b_{2l1/2}(t)|^2,
~~~\sigma_{33}^{0,0}(t)=\sum_{l,\bar l}|b_{21l\bar l}(t)|^2,
\nonumber\\
&&\sigma_{12}^{0,0}(t)=\sum_{l}b_{1l1/2}(t) \sb^*_{2l1/2}(t),~~~
\sigma_{00}^{1,0}(t)=\sum_{l,r}|b_{1l1/2r1/2}(t)|^2,~~~\ldots
\label{b2}
\end{eqnarray}
The diagonal density matrix elements, $\sigma_{jj}^{n,m}$, are the
probabilities of finding the system in one of the states shown in
Fig.~2 and the off-diagonal matrix elements (``coherencies'')
describe a linear superposition of these states. \vskip1cm
\begin{minipage}{13cm}
\begin{center}
\leavevmode \epsfxsize=14cm \epsffile{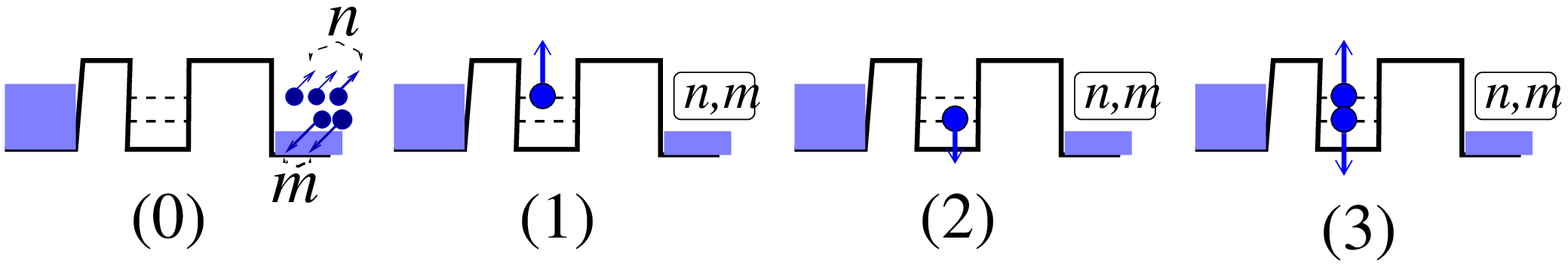}
\end{center}
{\begin{small} Fig.~2. Four available states of the quantum dot. The
indices $n,m$ denote the number of electrons with the spin
components $s'=\pm 1/2$ in the right reservoir.
\end{small}}
\end{minipage} \\ \\

It was demonstrated in Ref.\cite{g,gg} that the Schr\"odinger
equation for the entire system, $i\partial_t|\bPsi (t)\rangle
={\cal H}|\bPsi (t)\rangle$, can be reduced to Bloch-type rate
equations describing the reduced density-matrix
$\sigma_{jj'}^{n,m}(t)$. This reduction takes place after partial
tracing over the reservoir states. It becomes exact in the limit
of large bias without the explicit use of any Markov-type or weak
coupling approximations\cite{fn1}. In the general case these
equations are\cite {gg}
\begin{eqnarray}
&&\dot\sigma_{jj'}=i(E_{j' }-E_j)\sigma_{jj'} +
i\left (\sum_{k}\sigma_{jk}\tilde\Omega_{k\to j'}
-\sum_{k}\tilde\Omega_{j\to k}
\sigma_{kj'}\right )
\nonumber\\
&&-{\sum_{k,k'}}{\cal P}_2
\pi\rho(\sigma_{jk}\Omega_{k\to k'}\Omega_{k'\to j'}
+\sigma_{kj'}\Omega_{k\to k'}\Omega_{k'\to j})
 +\sum_{k,k'}{\cal P}_2
\pi\rho\,(\Omega_{k\to j}\Omega_{k'\to j'}+
\Omega_{k\to j'}\Omega_{k'\to j})
\sigma_{kk'} \, ,\nonumber\\
\label{b3}
\end{eqnarray}
\noindent (for simplicity we have omitted the indices $m$ and $n$,
which, however, can be easily restored from the conservation of
the total number of electrons). Here $\Omega_{k\to k'}$ denotes
the single-electron hopping amplitude that generates the $k\to k'$
transition. We distinguish between the amplitudes $\tilde\Omega$
describing single-electron hopping among isolated states and
$\Omega$ describing transitions among isolated and continuum
states. The latter can generate transitions between the isolated
states of the system, but only indirectly, via two consecutive
jumps of an electron, into and out of the {\em continuum}
reservoir states (with the density of states $\rho$). These
transitions are represented by the third and the fourth terms of
Eq.~(\ref{b3}). The third term describes the transitions ($k\to k'
\to j$) or ($k\to k' \to j'$), which cannot change the number of
electrons ($n,m$) in the collector. The fourth term describes the
transitions ($k\to j$ and $k' \to j'$) or ($k\to j'$ and $k' \to
j$) which increase the number of electrons in the collector by
one. These two terms of Eq.~(\ref{b3}) are analogues of the
``loss'' (negative) and the ``gain'' (positive) terms in the
classical rate equations, respectively. Yet, the sign of these
terms depends on the relative sign of the corresponding couplings
$\Omega$\cite{ieee}. In our case it is determined by the sign of
the spin-flip amplitude, Eq.~(\ref{a33}). In addition, there is a
(permutation) operator, ${\cal P}_2=\pm 1$, due to
anti-commutation of the fermions operators, $a_{1,2}^\dagger$ in
Eq.~(\ref{b1}) (See also\cite{g}). The prefactor ${\cal P}_2=-1$
whenever the loss or the gain terms in Eq.~(\ref{b3}) are
generated by a two-particle state of the dot. Otherwise ${\cal
P}_2=1$.

\section{Non-interacting electrons.}

Consider first the case of no electron repulsion inside the dot,
$U_C=0$. (In fact, the results would be the same for
$U_C\ll\mu_L-E_1$, assuming the couplings $\Omega$ are independent
of energy). As in the previous section we choose the initial
(``vacuum'') state corresponding to the polarized electrons in the
left reservoir, $s=1/2$ (Fig.~1). In this case all four
configurations shown in Fig.~3 contribute to Eqs.~(\ref{b3}).
Taking into account that there is no direct coupling between the
states, $E_{1,2}$, i.e. $\tilde\Omega =0$, one obtains the
following Bloch-type rate equations for the density matrix
$\sigma_{jj'}^{n,m}(t)$
\begin{mathletters}
\label{b4}
\begin{eqnarray}
&&\dot\sigma_{00}^{n,m}=-\Gamma_L\sigma_{00}^{n,m}
+\Gamma_R^{(1)}(\sigma_{11}^{n-1,m}+\sigma_{22}^{n,m-1})
+\Gamma_R^{(2)}(\sigma_{11}^{n,m-1}+\sigma_{22}^{n-1,m})
+\Gamma_L^{(2)}\sigma_{11}^{n,m} +\Gamma_L^{(1)}\sigma_{22}^{n,m}
\nonumber\\[5pt]
&&~~~~~~~~~~~~~~~~~~~~~~~~+\Gamma_L^{(12)}(\sigma_{12}^{n,m}
+\sigma_{21}^{n,m})-\Gamma_R^{(12)}
(\sigma_{12}^{n-1,m}+\sigma_{21}^{n-1,m}-
\sigma_{12}^{n,m-1}-\sigma_{21}^{n,m-1})
\label{b4a}\\[5pt]
&&\dot\sigma_{11}^{n,m}=-\left(\Gamma_R
  +2\Gamma_L^{(2)}\right)\sigma_{11}^{n,m}
+\Gamma_L^{(1)}(\sigma_{00}^{n,m}+\sigma_{33}^{n,m})
-\Gamma_L^{(12)}(\sigma_{12}^{n,m}+\sigma_{21}^{n,m})\nonumber\\[5pt]
&&~~~~~~~~~~~~~~~~~~~~~~~~~~~~~~~~~~~~~~~~~~~~~~~~~~~~~~~~~~~~~~~~~~~~~~~~~~~~
+\Gamma_R^{(2)}\sigma_{33}^{n-1,m}
+\Gamma_R^{(1)}\sigma_{33}^{n,m-1}
\label{b4b}\\[5pt]
&&\dot\sigma_{22}^{n,m}=-\left(\Gamma_R
  +2\Gamma_L^{(1)}\right)\sigma_{22}^{n,m}
+\Gamma_L^{(2)}(\sigma_{00}^{n,m}+\sigma_{33}^{n,m})
-\Gamma_L^{(12)}(\sigma_{12}^{n,m}+\sigma_{21}^{n,m})
\nonumber\\[5pt]
&&~~~~~~~~~~~~~~~~~~~~~~~~~~~~~~~~~~~~~~~~~~~~~~~~~~~~~~~~~~~~~~~~~~~~~~~~~~~~
+\Gamma_R^{(1)}\sigma_{33}^{n-1,m}
+\Gamma_R^{(2)}\sigma_{33}^{n,m-1}
\label{b4c}\\[5pt]
&&\dot\sigma_{33}^{n,m}=-(2\Gamma_R
  +\Gamma_L)\sigma_{33}^{n,m}
  +\Gamma_L^{(2)}\sigma_{11}^{n,m}
    +\Gamma_L^{(1)}\sigma_{22}^{n,m}
+\Gamma_L^{(12)}(\sigma_{12}^{n,m} +\sigma_{21}^{n,m})
\label{b4d}\\[5pt]
&&\dot\sigma_{12}^{n,m}=-(i\epsilon +\Gamma )\sigma_{12}^{n,m}
-\Gamma_L^{(12)}(\sigma_{00}^{n,m}+\sigma_{11}^{n,m}
+\sigma_{22}^{n,m}+\sigma_{33}^{n,m})
+\Gamma_R^{(12)}(\sigma_{33}^{n-1,m}-\sigma_{33}^{n,m-1})\, ,
\label{b4e}
\end{eqnarray}
\end{mathletters}
where $\Gamma =\Gamma_{L}+\Gamma_{R}$ and
$\Gamma_{L,R}^{(1)}=\Gamma_{L,R}\cos^2(\theta_{L,R}/2)$,
$\Gamma_{L,R}^{(2)}=\Gamma_{L,R}\sin^2(\theta_{L,R}/2)$,
$\Gamma_{L,R}^{(12)}=\Gamma_{L,R}\sin (\theta_{L,R}/2)\cos
(\theta_{L,R}/2)$ are the partial tunneling widths of the levels
$E_{1,2}$.

We now trace the origin of each term in these equations taking as an
example Eq.~(\ref{b4b}), corresponding to $j=j'=1$ in Eq.(\ref{b3}).
The first term in this equation is a ``loss'' term generated by the
transitions $1\to 0\to 1$ and $1\to 3\to 1$ in Eq.~(\ref{b3}).
corresponding to the following processes: (a) an electron at the
level $E_1$ (Fig.~3(1)) tunnels to the right reservoir and back to
the same state, with the rate $\Gamma_R$; (b) the same electron
tunnels to the available continuum states of the left reservoir and
back to the level $E_1$, with the rate
$\Gamma_L^{(2)}=\Gamma_L\sin^2(\theta_L/2)$. This can proceed only
via spin-flip, since the spin-up states in the left reservoir are
occupied; (c) an electron from occupied states of the left reservoir
tunnels to the unoccupied level $E_2$ of the dot and then back to
the same state of the left reservoir, with the rate
$\Gamma_L^{(2)}$.

The second term in Eq.~(\ref{b4b}) is a ``gain'' term generated by
the transitions $0\to 1, 0\to 1$ and $3\to 1, 3\to 1$ of an
electron from the left reservoir to the level $E_1$ and from the
level $E_2$ to the unoccupied (spin-down) continuum states of the
left reservoir.

The third, ``loss'', term in Eq.~(\ref{b4b}) is generated by the
transitions $2\to 0\to 1$ and $2\to 3\to 1$ via the left
reservoir. These transitions involve the following processes: (a)
an electron at the level $E_2$ (Fig.~3(2)) tunnels to an
unoccupied, spin-down state of the left reservoir, and then makes
a spin-flip transition to the state $E_1$ of the dot. The rate of
this process is $(1/2)\Gamma_L\sin (\theta_L) \cos (\theta_L/2)$,
as follows from Eq.~(\ref{b3}); (b) an electron from one of the
occupied states of the left reservoir tunnels to the state $E_1$
with the corresponding amplitude $\Omega_L\cos (\theta_L/2)$. Then
an electron with energy $E_2$ tunnels to the vacant state of the
left reservoir with the spin-flip amplitude $-\Omega_L\sin
(\theta_L/2)$. Since this transition proceeds via the two-electron
state of the dot, the corresponding permutation prefactor, ${\cal
P}_2=-1$. As a result, the rate of this (loss) process is
$(1/2)\Gamma_L\sin (\theta_L)$. Similar transitions via the right
reservoir cancel. Indeed the electron from energy level $E_2$ can
reach the level $E_1$ by two ways: the first through the spin-flip
hopping to the right reservoir and then to the level $E_1$ with no
spin-flip, and the second, without spin-flip to the right
reservoir, and then to the level $E_1$ with the spin flip. These
two amplitudes are of the opposite sign.

The last two terms of Eq.~(\ref{b4b}) are ``gain'' terms generated
by the transitions $3\to 1, 3\to 1$ of an electron from the state in
Fig.~3(3) to the spin-up or spin-down states of the right reservoir.
The number of electrons in the right reservoir increases by one.

\subsection{Resonant current in the collector}

Using Eqs.~(\ref{b4}) we can easily obtain the spin-up and
spin-down currents, $I_{1/2}(t)=\sum_{n,m}n\dot P_{n,m}(t)$ and
$I_{-1/2}(t)=\sum_{n,m}m\dot P_{n,m}(t)$, where
$P_{n,m}(t)=\sum_{j=0}^{j=3}\sigma^{n,m}_{jj}$ is the probability
of finding $n$ electrons with spin up and $m$ electrons with spin
down in the right reservoir. One finds
\begin{equation}
I_{\pm 1/2}(t) =\Gamma_R\left[{1\pm\cos\theta_R\over2}\sigma_{11}(t)
  +{1\mp\cos\theta_R\over2}\sigma_{22}(t)+\sigma_{33}(t)
  \mp{\sin\theta_R\over2}(\sigma_{12}(t)+\sigma_{21}(t))\right ]
\label{b5}
\end{equation}
where $\sigma_{jj'}(t)=\sum_{n,m}\sigma_{jj'}^{n,m}(t)$. The latter
can be obtained from the following matrix equation
\begin{equation}
  \dot X(t)+BX(t)=0\, ,
  \label{b6}
\end{equation}
obtained by the summation of Eqs.~(\ref{b4}) over $n,m$. Here
$X=\{\sigma_{00},\sigma_{11},\sigma_{22},
\sigma_{33},\sigma_{12},\sigma_{21}\}$ and $B$ is the
corresponding $6\times 6$ matrix,
  \begin{equation}
 B= \pmatrix{\Gamma_L&-\Gamma_L^{(2)}-\Gamma_R&
    -\Gamma_L^{(2)}-\Gamma_R&0&-\Gamma_L^{(12)}&
    -\Gamma_L^{(12)}\cr
    -\Gamma_L^{(1)}&\Gamma_R+2\Gamma_L^{(2)}&
    0&-\Gamma_L^{(1)}-\Gamma_R&\Gamma_L^{(12)}&
     \Gamma_L^{(12)}\cr
 -\Gamma_L^{(2)}&0&\Gamma_R+2\Gamma_L^{(1)}&
    -\Gamma_L^{(2)}-\Gamma_R&\Gamma_L^{(12)}&
     \Gamma_L^{(12)}\cr
     0&-\Gamma_L^{(2)}&-\Gamma_L^{(1)}&
     \Gamma_L+2\Gamma_R&-\Gamma_L^{(12)}&
     -\Gamma_L^{(12)}\cr
     \Gamma_L^{(12)}&\Gamma_L^{(12)}
     &\Gamma_L^{(12)}&\Gamma_L^{(12)}&
     i\epsilon+\Gamma&0\cr
    \Gamma_L^{(12)}&\Gamma_L^{(12)}
     &\Gamma_L^{(12)}&\Gamma_L^{(12)}&0&
     -i\epsilon+\Gamma}
   \label{bb6}
\end{equation}

Solving Eqs.~({\ref{b6}) and substituting the result into
Eqs.~(\ref{b5}) we find the following simple expressions for the
average polarized current:
\begin{eqnarray}
  && I_{\pm 1/2}(t)={\Gamma_L\Gamma_R\over 2\Gamma}(1\pm\cos \theta_L\cos
  \theta_R)
  (1-e^{-\Gamma t})
\nonumber\\
&&~~~~~~~~~~~~~~~~~~~~~~~~~~~~~~~~~ \pm {\Gamma_L\Gamma_R\Gamma\sin
\theta_L\sin \theta_R\over 2(\epsilon^2+\Gamma^2)} \left
[1-e^{-\Gamma t}\cos (\epsilon t) +e^{-\Gamma
t}{\epsilon\over\Gamma} \sin (\epsilon t)\right ]\, . \label{a13}
\end{eqnarray}
The same result can be obtained in the framework of a single
electron approach, valid for the noninteracting case. (See
Appendix A.) As expected, the polarized resonant current displays
damped oscillations. An example of these oscillations in
$I_{1/2}(t)$ is shown in Fig.~3. \vskip1cm
\begin{minipage}{13cm}
\begin{center}
\leavevmode \epsfxsize=8cm \epsffile{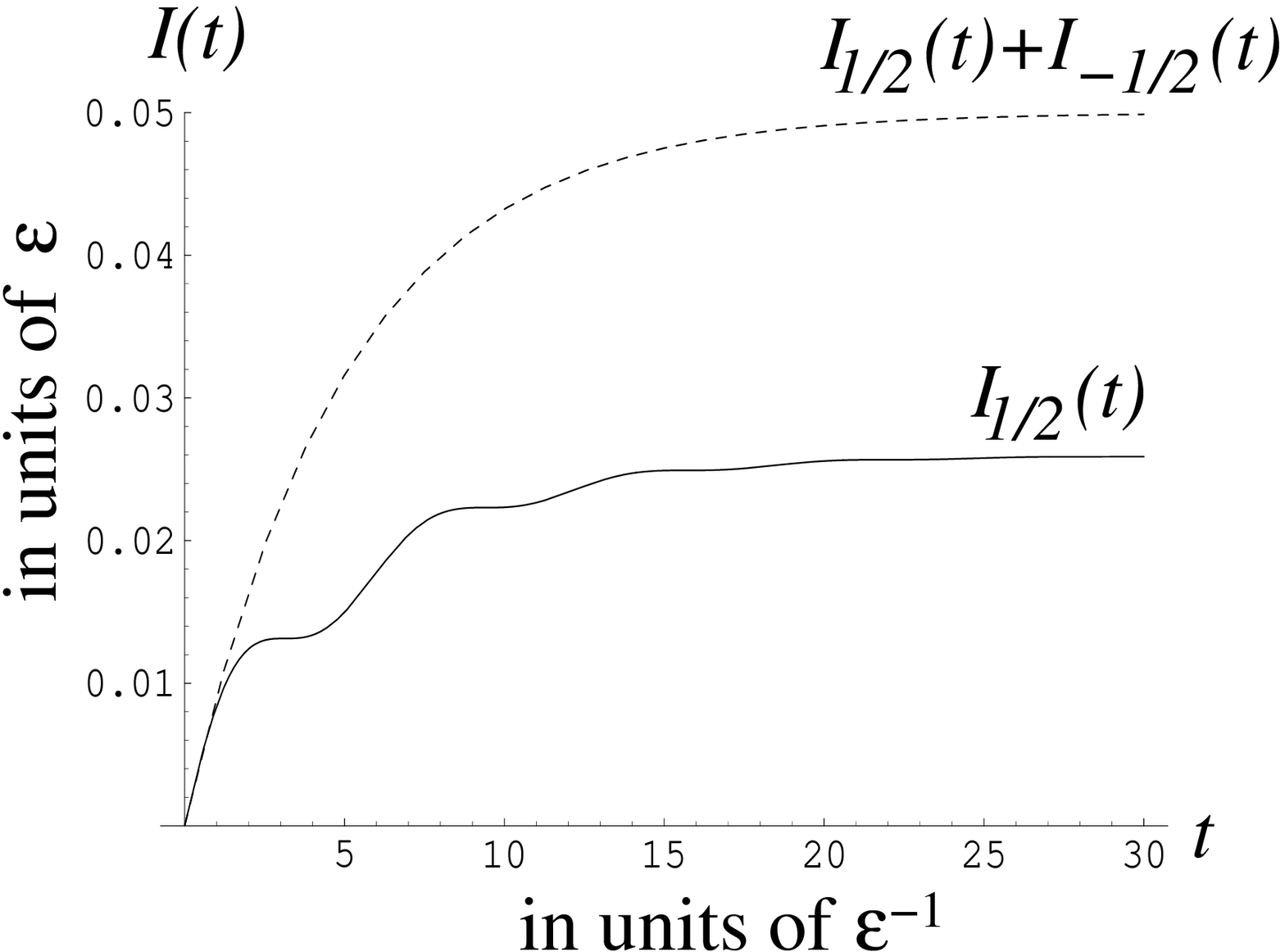}
\end{center}
{\begin{small} Fig.~3. Spin-up and total resonant currents through
the Zeeman doublet as a function of time for
$\theta_L=\theta_R=\pi/2$ and $\Gamma_L=\Gamma_R=0.1\epsilon$.
\end{small}}
\end{minipage} \\ \\
These oscillations, however, disappear in the total collector
current,
\begin{equation}
I(t)=
I_{1/2}(t)+I_{-1/2}(t)={\Gamma_L\Gamma_R\over\Gamma}(1-e^{-\Gamma
t})\, , \label{aa16}
\end{equation}
even though electrons in the emitter are polarized.

\subsection{Noise spectrum of the total collector current}

Now we evaluate the noise-spectrum of the total current, represented
by a sum of the spin-up and spin-down currents in the final state,
Eq.~(\ref{aa16}). We introduce the density matrix
$\sigma^N_{jj'}(t)= \sum_n\sigma^{n,N-n}(t)$, obtained from
Eqs.~(\ref{b4}), where $N$ denotes the total number of electrons
which have arrived at the right reservoir by time $t$. In order to
calculate the shot-noise spectrum we use the McDonald
formula\cite{mcdon}
\begin{equation}
S(\omega) = 2e^2\omega \int_0^\infty dt
\sin (\omega t) {d\over dt}\sum_NN^2P_N(t)]\, ,
\label{b7}
\end{equation}
where $P_N(t) =\sum_{j=0}^{j=3}\sigma_{jj}^N(t)$. One easily finds
from Eqs.~(\ref{b4}) that
\begin{equation}
  \sum_NN^2\dot P_N(t)=\Gamma_R\sum_N(2N+1)\left[
   \sigma^N_{11}(t)+\sigma^N_{22}(t)+2\sigma^N_{33}(t)\right]\, .
\label{b9}
\end{equation}

Substituting Eq.~(\ref{b9}) into the McDonald formula, Eq.~(\ref{b7}), we
finally obtain
\begin{equation}
  S(\omega )= 2e^2\omega\Gamma_R {\mbox{Im}}\, \left[
  Z_{11}(\omega )+Z_{22}(\omega )+2Z_{33}(\omega )\right]\, ,
\label{b10}
\end{equation}
where $Z(\omega )$ is a $6$-vector,
$Z=\{Z_{00},Z_{11},Z_{22},Z_{33},Z_{12}Z_{21}\}$, defined as
\begin{equation}
Z_{ij}(\omega) = \int_0^\infty \sum_N(2N+1)\sigma_{ij}^N(t)\exp
(i\omega t)dt\, .
\label{b11}
\end{equation}
One can find $Z_{ij}(\omega)$ directly from Eqs.~(\ref{b4})
by performing the corresponding summation over $N$. As
a result one obtains
\begin{equation}
  (B-i\omega I)Z(\omega) =\bar X+2\Gamma_R\bar Y(\omega )\, .
\label{b12}
\end{equation}
Here $B$ is given by Eq.~(\ref{bb6}) and  $I$ is the unit matrix.
The 6-vector $\bar X$ corresponds to the stationary solution of
Eqs.~(\ref{b6}), $\bar X=X(t\to\infty )$ and $\bar Y(\omega
)=\{Y_{11}+Y_{22},Y_{33},Y_{33},0,0,0\}$ where $ Y(\omega
)=\{Y_{00},Y_{11},Y_{22},Y_{33},Y_{33},Y_{12},Y_{21}\}$ is given
by the equation
\begin{equation}
  (B-i\omega I)Y(\omega) =\bar X
\label{b13}
\end{equation}

Using Eq.~(\ref{b12}) we calculate the ratio
of the shot-noise power spectrum to the Schottky noise,
$S(\omega )/2eI$ (Fano factor), where
$I=I(t\to\infty )=\Gamma_L\Gamma_R/\Gamma_t$, Eq.(\ref{aa16}).
In particular, the result has a simple analytical form for a symmetric
dot, $\Gamma_L= \Gamma_R=\Gamma$. We find
\begin{equation}
{S(\omega )\over 2eI}={2\Gamma^2+\omega^2\over 4\Gamma^2+\omega^2}+
{\Gamma^2\epsilon^2\sin^2\theta_L\over  (4\Gamma^2+\omega^2)(4\Gamma^2+\omega^2)}\, .
\label{bb13}
\end{equation}
As expected the shot-noise spectrum does not display any peak or dip
at frequencies corresponding to the Zeeman splitting, since the
interference effects are canceled in the total collector  current.
Yet the noise spectrum depends on the initial polarization of
incoming electrons ($\theta_L$), whereas the total collector current
does not (see Eq.~(\ref{aa16})). If electrons are initially
polarized along the magnetic field inside the dot ({\bf \={n}}), the
Fano factor is the same as in the case of resonant tunneling through
a single level\cite{chen}. With increasing $\theta_L$, however, the
current flows through both levels of the Zeeman doublet. This leads
to an additional contribution to the shot noise, described by the
second term of Eq.~(\ref{bb13}).

\subsection{Ferromagnetic reservoirs}
Let us consider ferromagnetic reservoirs polarized along {\bf n}
and {\bf {n'}} directions, Fig.~1. In this case the rate equations
(\ref{b4}) have to be modified since there are no available
spin-down states in the left and right reservoirs. One easily
obtains the following rate equations for the density matrix
$\sigma_{jj'}^n(t)$, where $n$ denotes the number of electron,
arriving at the collector before time $t$:
\begin{mathletters}
\label{b16}
\begin{eqnarray}
\dot\sigma_{00}^{n}&=&-\Gamma_L\sigma_{00}^{n}
+\Gamma_R^{(1)}\sigma_{11}^{n-1} +\Gamma_R^{(2)}\sigma_{22}^{n-1}
-\Gamma_R^{(12)} (\sigma_{12}^{n-1}+\sigma_{21}^{n-1})
\label{b16a}\\[5pt]
\dot\sigma_{11}^{n}&=&-\left(\Gamma_L^{(2)}
  +\Gamma_R^{(1)}\right)\sigma_{11}^{n}
+\Gamma_L^{(1)}\sigma_{00}^{n}
-{1\over2}(\Gamma_L^{(12)}-\Gamma_R^{(12)})(\sigma_{12}^{n}+\sigma_{21}^{n})
+\Gamma_R^{(2)}\sigma_{33}^{n-1}
\label{b16b}\\[5pt]
\dot\sigma_{22}^{n}&=&-\left(\Gamma_L^{(1)}
  +\Gamma_R^{(2)}\right)\sigma_{22}^{n}
+\Gamma_L^{(2)}\sigma_{00}^{n}
-{1\over2}(\Gamma_L^{(12)}-\Gamma_R^{(12)})(\sigma_{12}^{n}+\sigma_{21}^{n})
+\Gamma_R^{(1)}\sigma_{33}^{n-1}
\label{b16c}\\[5pt]
\dot\sigma_{33}^{n}&=&-\Gamma_R\sigma_{33}^{n}
  +\Gamma_L^{(2)}\sigma_{11}^{n}
    +\Gamma_L^{(1)}\sigma_{22}^{n}
+\Gamma_L^{(12)}(\sigma_{12}^{n} +\sigma_{21}^{n})
\label{b16d}\\[5pt]
\dot\sigma_{12}^{n}&=&-\left(i\epsilon
+{1\over2}\Gamma\right)\sigma_{12}^{n}
-\Gamma_L^{(12)}\sigma_{00}^{n}-{1\over2}(\Gamma_L^{(12)}-\Gamma_R^{(12)})
(\sigma_{11}^{n}+\sigma_{22}^{n}) +\Gamma_R^{(12)}\sigma_{33}^{n-1}
\label{b16e}
\end{eqnarray}
\end{mathletters}

Using these equations we first evaluate the average current,
$I(t)\equiv I_{1/2}(t)$ given by Eq.~(\ref{b5}) with
$\sigma_{jj'}(t)=\sum_n\sigma_{jj'}^n(t)$. The latter quantities
are obtained from a summation of Eqs.~(\ref{b16}) over $n$. As a
result Eqs.~(\ref{b16}) are reduced to the matrix equation
(\ref{b6}), where $B$ is the corresponding  $6\times 6$ matrix of
the coefficients of Eqs.~(\ref{b16}). Solving this equation we
find the average current $I(t)$.  For instance, in the case of
$\theta_L=\theta_R$ one finds for the stationary current,
$I=I(\infty ) =\Gamma_L\Gamma_R/(\Gamma_L+\Gamma_R)$. One obtains
the same expression for the resonant tunneling of unpolarized
electrons through a single level.

The time dependence of the average current, $I(t)$, is displayed
in Fig.~4 for symmetric and asymmetric dots,
$\Gamma_L=\Gamma_R=0.1\epsilon$ and $\Gamma_L=\epsilon$,
$\Gamma_R=0.1\epsilon$, respectively. Comparing with Fig.~2 one
finds that the oscillations in the average current are more
pronounced in the case of ferromagnetic reservoirs. This can be
anticipated since the corresponding spin-flip transitions via the
spin-down states of the reservoirs do not exist. We recall that
precisely these transitions resulted in the cancelation of the
interference effects in the  previous case.
 \vskip1cm
\begin{minipage}{13cm}
\begin{center}
\leavevmode
\epsfxsize=8cm
\epsffile{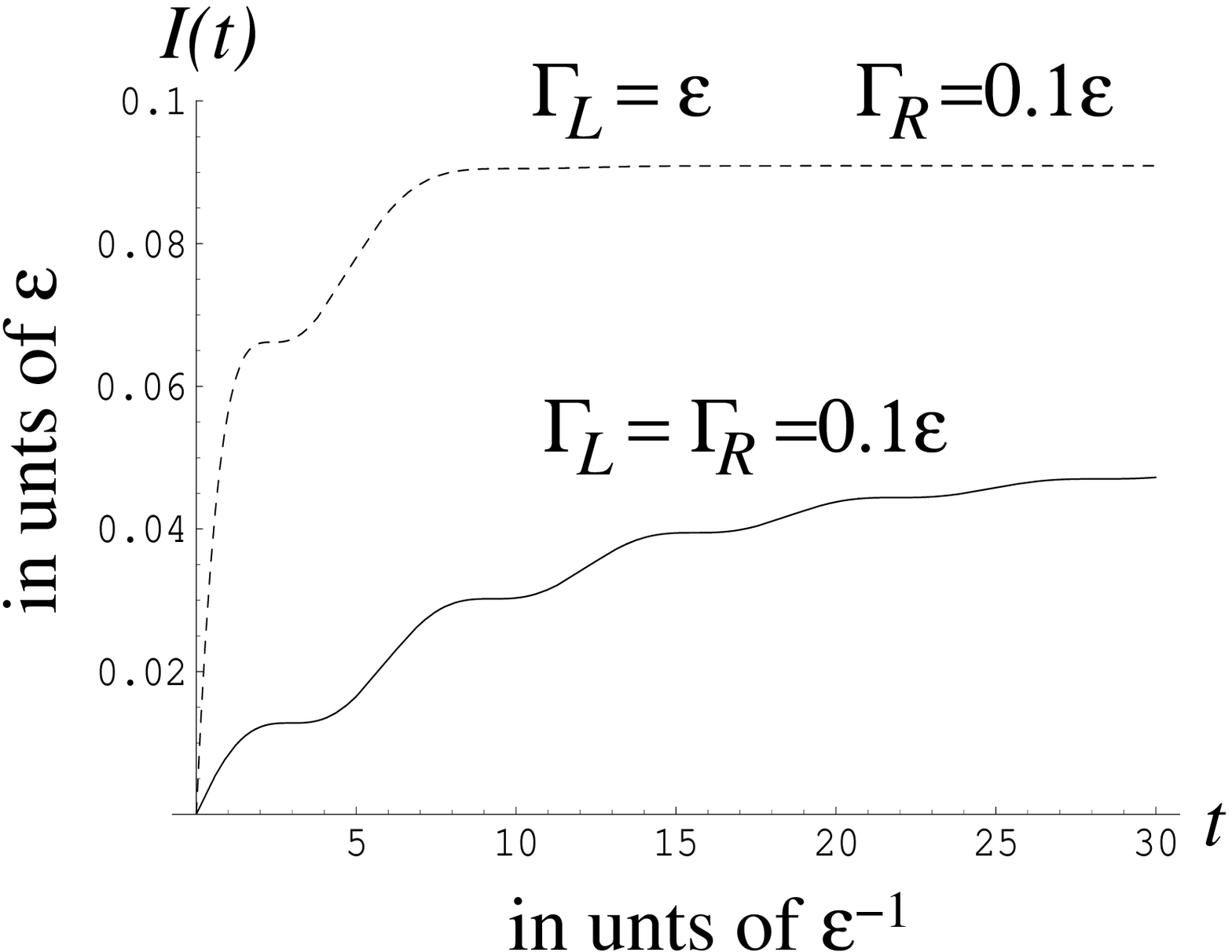}
\end{center}
{\begin{small} Fig.~4. The polarized resonant current through the
Zeeman doublet with ferromagnetic reservoirs and
$\theta_L=\theta_R=\pi/2$. The solid line corresponds to
$\Gamma_L=\Gamma_R=0.1\epsilon$ and the dashed line to
$\Gamma_L=\epsilon$ and $\Gamma_R=0.1\epsilon$.
\end{small}}
\end{minipage} \\ \\

Now  we can evaluate the shot-noise spectrum, $S(\omega )$ using
the McDonald formula. One obtains from Eqs.~(\ref{b7}) and
(\ref{b16})
\begin{eqnarray}
  S(\omega )&=& 2e^2\omega{\mbox{Im}}\, \left\{
  \Gamma_R^{(1)} Z_{11}(\omega )
  +\Gamma_R^{(2)}Z_{22}(\omega )+\Gamma_RZ_{33}(\omega )
-\Gamma_R^{(12)}[Z_{12}(\omega )+Z_{21}(\omega )]\right\}\, ,
\label{b17}
\end{eqnarray}
where $Z(\omega )$ is given by Eq.~(\ref{b12}) with the matrix $B$
corresponding to Eqs.~({\ref{b16}) and $\bar Y =\{ \bar Y_{00},\bar
Y_{11},\bar Y_{22},0,\bar Y_{12},\bar Y_{21}\}$. Here $\bar
Y_{00}=\cos^2{\theta_R\over2}Y_{11}
    +\sin^2{\theta_R\over2}Y_{22}
    -{\sin\theta_R\over2}(Y_{12}+Y_{21})$, $\bar Y_{11}=
    \sin^2{\theta_R\over2}Y_{33}$, $\bar Y_{22}=\cos^2{\theta_R\over2}Y_{33}$,
    and $\bar Y_{12}=\bar Y_{21}= {\sin\theta_R\over2}Y_{33}$, while
 $Y_{jj'}=Y_{jj'}(\omega )$ are given by Eq.~(\ref{b13}).

The corresponding Fano factor is shown in Fig.~5 for the same
parameters as in Fig.~4. It clearly displays a dip at the Zeeman
frequency for a symmetric dot. It reflects the damped oscillations
in the average current, shown in Fig.~4. The dip, however, almost
disappears for an asymmetric dot with large $\Gamma_L$.

\vskip1cm
\begin{minipage}{13cm}
\begin{center}
\leavevmode
\epsfxsize=11cm
\epsffile{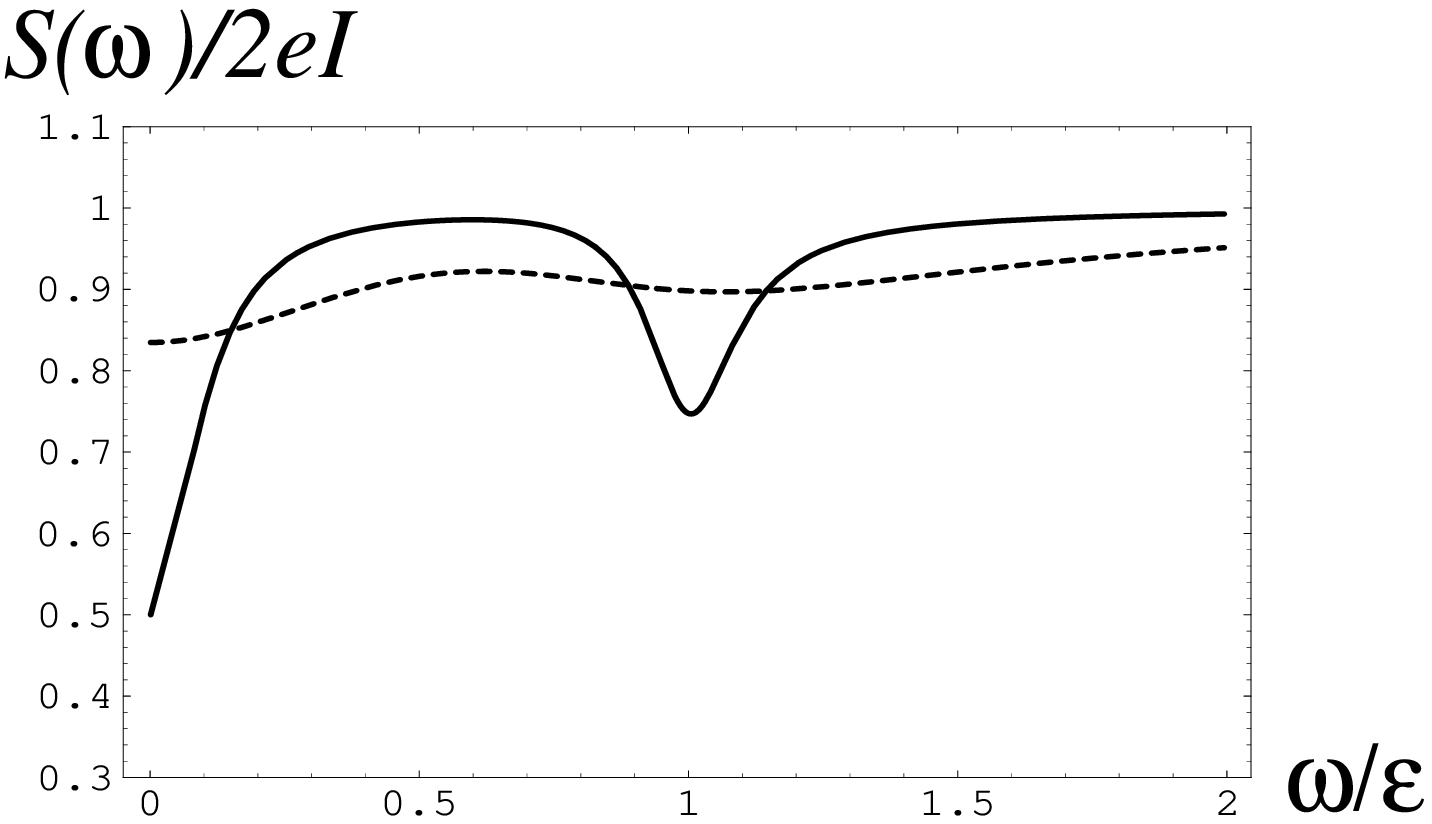}
\end{center}
{\begin{small} Fig.~5. The Fano factor versus $\omega $ for a
polarized electron current with ferromagnetic reservoirs and
$\theta_L,\theta_R=\pi/2$. The solid line corresponds to
$\Gamma_L=\Gamma_R=0.1\epsilon$ and the dashed line to
$\Gamma_L=\epsilon$ and $\Gamma_R=0.1\epsilon$.
\end{small}}
\end{minipage} \\ \\

\section{Coulomb blockade}

We now introduce strong Coulomb repulsion inside the dot,
$U_C\gg\mu_L-E_1$, so that the state (3) in Fig.~3 is not
available. As a result the corresponding rate equations have an
even simpler form than those found for non-interacting electrons.
Consider again the case of ferromagnetic reservoirs, where the
quantum interference effects are most pronounced. The
corresponding rate equations for the case of Coulomb blockade can
be obtained from Eqs.~(\ref{b16}) for non-interacting electrons,
by eliminating configurations with two electrons in the dot. In
the following we consider separately the electron current in the
right and in the left reservoirs.

\subsection{Collector current}
The electrical current in the right reservoir and its power spectrum
are obtained from the following rate equation
\begin{mathletters}
\label{c1}
\begin{eqnarray}
\dot\sigma_{00}^{n}&=&-\Gamma_L\sigma_{00}^{n}
+\Gamma_R^{(1)}\sigma_{11}^{n-1} +\Gamma_R^{(2)}\sigma_{22}^{n-1}
-\Gamma_R^{(12)} (\sigma_{12}^{n-1}+\sigma_{21}^{n-1})
\label{c1a}\\[5pt]
\dot\sigma_{11}^{n}&=&-\Gamma_R^{(1)}\sigma_{11}^{n}
+\Gamma_L^{(1)}\sigma_{00}^{n}
+{\Gamma_R^{(12)}\over2}(\sigma_{12}^{n}+\sigma_{21}^{n})
\label{c1b}\\[5pt]
\dot\sigma_{22}^{n}&=&-\Gamma_R^{(2)}\sigma_{22}^{n}
+\Gamma_L^{(2)}\sigma_{00}^{n}
+{\Gamma_R^{(12)}\over2}(\sigma_{12}^{n}+\sigma_{21}^{n})
\label{c1c}\\[5pt]
\dot\sigma_{12}^{n}&=&-\left (i\epsilon +{\Gamma_R\over2}\right
)\sigma_{12}^{n} -\Gamma_L^{(12)}\sigma_{00}^{n}
+{\Gamma_R^{(12)}\over2}(\sigma_{11}^{n}+\sigma_{22}^{n})
\label{c1e}
\end{eqnarray}
\end{mathletters}

Using these equations one finds for the average (polarized) current
in the collector
\begin{equation}
I_R(t) =\Gamma_R^{(1)}\sigma_{11}(t)
  +\Gamma_R^{(2)}\sigma_{22}(t)
  -\Gamma_R^{(12)}[\sigma_{12}(t)+\sigma_{21}(t)]
\label{c5}
\end{equation}
where the $\sigma_{jj'}(t)=\sum_{n,m}\sigma_{jj'}^{n,m}(t)$ are
obtained from Eq.~(\ref{b6}) for
$X=\{\sigma_{00},\sigma_{11},\sigma_{22},
\sigma_{12},\sigma_{21}\}$, and $B$ is the $5\times 5$ matrix
obtained from the coefficients of Eqs.~(\ref{c1}). Solving such a
modified Eq.~(\ref{b6}) for $\theta_L=\theta_R$, one finds for the
stationary current,
 \begin{equation}
I_R=I_R(\infty ) ={\Gamma_L\Gamma_R\over 2\Gamma_L+\Gamma_R}
\label{cc7}
\end{equation}
This expression shows an asymmetry with respect to the widths
$\Gamma_L$ and $\Gamma_R$, in contrast with the non-interacting
case. The reason is that an electron enters the dot from the left
reservoir with the rate $2\Gamma_L$. However, it leaves it with
the rate $\Gamma_R$, since the state with two levels of the dot
occupied is forbidden.

The shot-noise power spectrum for the collector current is given
by
\begin{equation}
  S_R(\omega )= 2e^2\omega {\mbox{Im}}\, \left\{
  \Gamma_R^{(1)}Z_{11}(\omega )
  +\Gamma_R^{(2)}Z_{22}(\omega )
-\Gamma_R^{(12)}[Z_{12}(\omega )+Z_{21}(\omega )]\right\}\, .
\label{c7}
\end{equation}
Here $Z_{ij}(\omega )$ are obtained from Eqs.~(\ref{b12}),(\ref{b13}),
where $\bar Y =\{ \bar Y_{00},0,0,0,0\}$ and
$\bar Y_{00}= \cos^2{\theta_R\over2}Y_{11}+
\sin^2{\theta_R\over2}Y_{22}-{\sin\theta_R\over2}(Y_{12}+Y_{21})$.

The results of our calculations of $S(\omega )$ for symmetric and
asymmetric quantum dots in the case of Coulomb blockade are shown in
Fig.~6. \vskip1cm
\begin{minipage}{13cm}
\begin{center}
\leavevmode
\epsfxsize=11cm
\epsffile{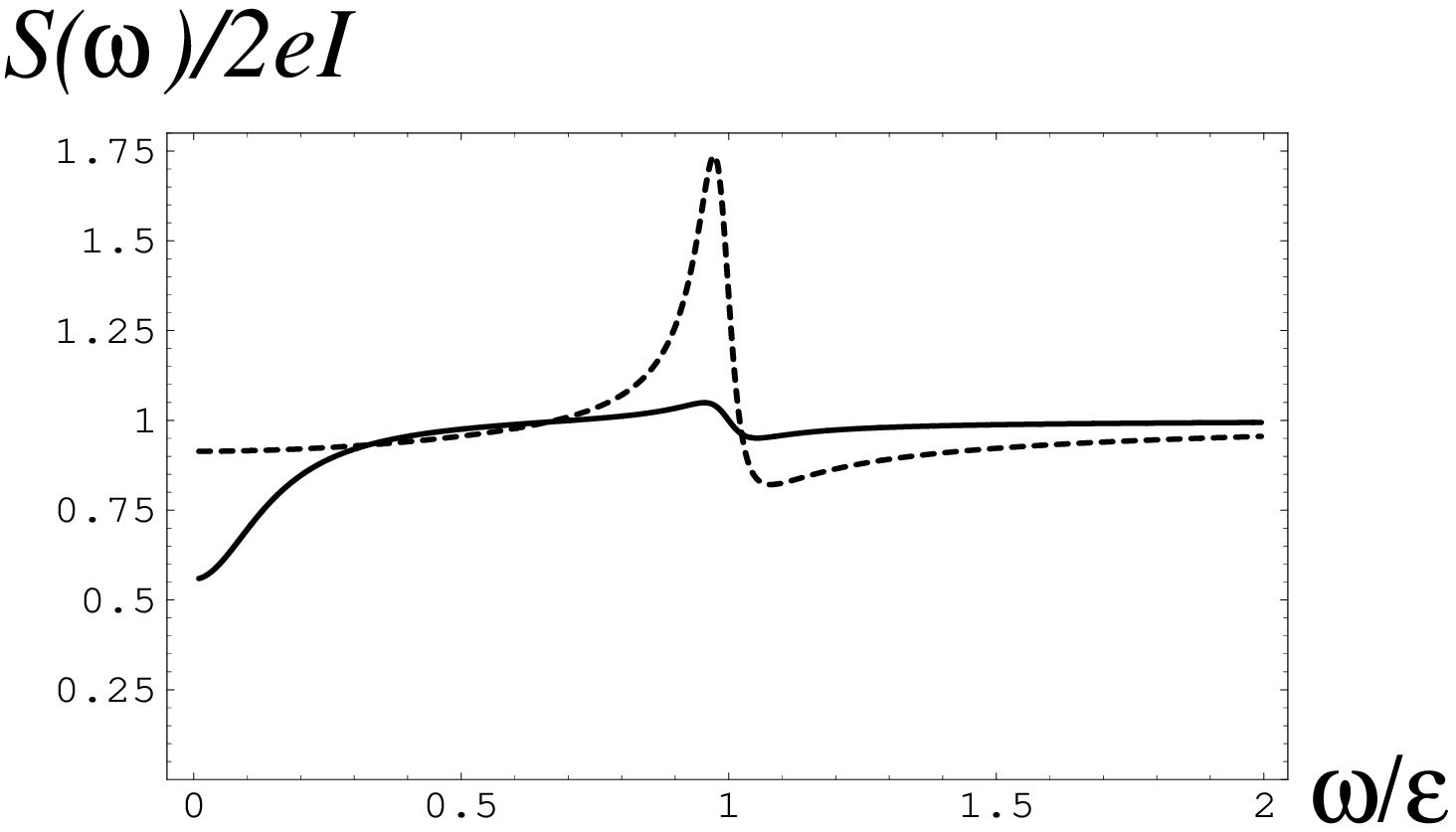}
\end{center}
{\begin{small} Fig.~6. The Fano factor versus $\omega $ for a
polarized collector current with ferromagnetic reservoirs and
Coulomb blockade and $\theta_L,\theta_R=\pi/2$. The solid line
corresponds to  $\Gamma_L=\Gamma_R=0.1\epsilon$ and the dashed
line to $\Gamma_L=\epsilon$ and $\Gamma_R=0.1\epsilon$.
\end{small}}
\end{minipage} \\ \\

\subsection{Emitter current}

We now consider the electric current and its power spectrum in the
left reservoir. These quantities are determined from the
density-matrix $\sigma_{jj'}^{p}$, where $p$ is the number of
electrons that left the emitter before time $t$, (the number of
holes in the left reservoir). The corresponding rate equations are
similar to Eqs.~(\ref{c1}). One finds
\begin{mathletters}
\label{c8}
\begin{eqnarray}
\dot\sigma_{00}^{p}&=&-\Gamma_L\sigma_{00}^{p}
+\Gamma_R^{(1)}\sigma_{11}^{p} +\Gamma_R^{(2)}\sigma_{22}^{p}
-\Gamma_R^{(12)}(\sigma_{12}^{p}+\sigma_{21}^{p})
\label{c8a}\\[5pt]
\dot\sigma_{11}^{p}&=&-\Gamma_R^{(1)}\sigma_{11}^{p}
+\Gamma_L^{(1)}\sigma_{00}^{p-1}
+{\Gamma_R^{(12)}\over2}(\sigma_{12}^{p}+\sigma_{21}^{p})
\label{c8b}\\[5pt]
\dot\sigma_{22}^{p}&=&-\Gamma_R^{(2)}\sigma_{22}^{p}
+\Gamma_L^{(2)}\sigma_{00}^{p-1}
+{\Gamma_R^{(12)}\over2}(\sigma_{12}^{p}+\sigma_{21}^{p})
\label{c8c}\\[5pt]
\dot\sigma_{12}^{p}&=&-\left (i\epsilon +{\Gamma_R\over2}\right
)\sigma_{12}^{p} -\Gamma_L^{(12)}\sigma_{00}^{p-1}
+{\Gamma_R^{(12)}\over2}(\sigma_{11}^{p}+\sigma_{22}^{p})
\label{c8e}
\end{eqnarray}
\end{mathletters}

The average emitter current in the left reservoir is given by
$I_L(t)=\Gamma_L\sigma_{00}(t)$, which differs from Eq.~(\ref{c5})
describing the collector current, $I_R(t)$. Yet, as expected, their
stationary values coincide, $I_L(\infty )=I_R(\infty )$,
Eq.~(\ref{cc7}).

The shot-noise power spectrum of the emitter current is given
\begin{equation}
S_L(\omega )=2e^2\omega\Gamma_L{\mbox{Im}}\, Z_{00}(\omega )
\label{c9}
\end{equation}
instead of Eq.~(\ref{c7}) for $S_R(\omega )$, where $Z_{ij}(\omega
)$ are obtained from Eqs.~(\ref{b12}),(\ref{b13}). Yet, $\bar
Y(\omega ) =\{ 0,\cos^2{\theta_L\over2},\sin^2{\theta_L\over2},
-{\sin\theta_L\over2},-{\sin\theta_L\over2}\}Y_{00}(\omega )$, in
contrast with the corresponding expression for $S_R(\omega )$.
Even though the expressions for $S_{L,R}(\omega )$ are quite
different, one finds that the shot-noise power of the emitter
current is the same as that in the collector current, $S_L(\omega
) =S_R(\omega )$.

\subsection{Circuit current}

In general the circuit current is given by $I_c(t)=\alpha
I_L(t)+\beta I_R(t)$, where the coefficients $\alpha,\beta$ with
$\alpha +\beta =1$ depend on the junction capacities\cite{but}.
Using charge conservation, $I_L=I_R+\dot{Q}$, where $Q$ is charge
in the dot, one finds
\begin{equation}
I_c(t)I_c(0)=\alpha I_L(t)I_L(0)+\beta I_R(t)I_R(0)-\alpha\beta \dot
Q(t)\dot Q(0)\, .\label{c10}
\end{equation}
Using this relation one finds a simple expression for the noise
spectrum of the circuit current\cite{moz,agu}
\begin{equation}
S_c(\omega )=\alpha S_L(\omega )+\beta S_R(\omega
)-\alpha\beta\omega^2S_Q(\omega )\, .\label{c11}
\end{equation}
where $S_Q(\omega )$ is Fourier transform of the charge correlation
function. This quantity can be obtained straightforwardly from the
matrix equation (\ref{b13}), where $\bar X$ is the 5-vector
$\{0,\sigma_{11}(\infty ),\sigma_{22}(\infty ),0,0\}$. Then
$S_Q(\omega )=4$Re$[Y_{11}(\omega )+Y_{22}(\omega )]$.

The results of our calculations of $S_c(\omega )$ for $\alpha =\beta
=1/2$ are shown in Fig.~7. \vskip1cm
\begin{minipage}{13cm}
\begin{center}
\leavevmode \epsfxsize=11cm \epsffile{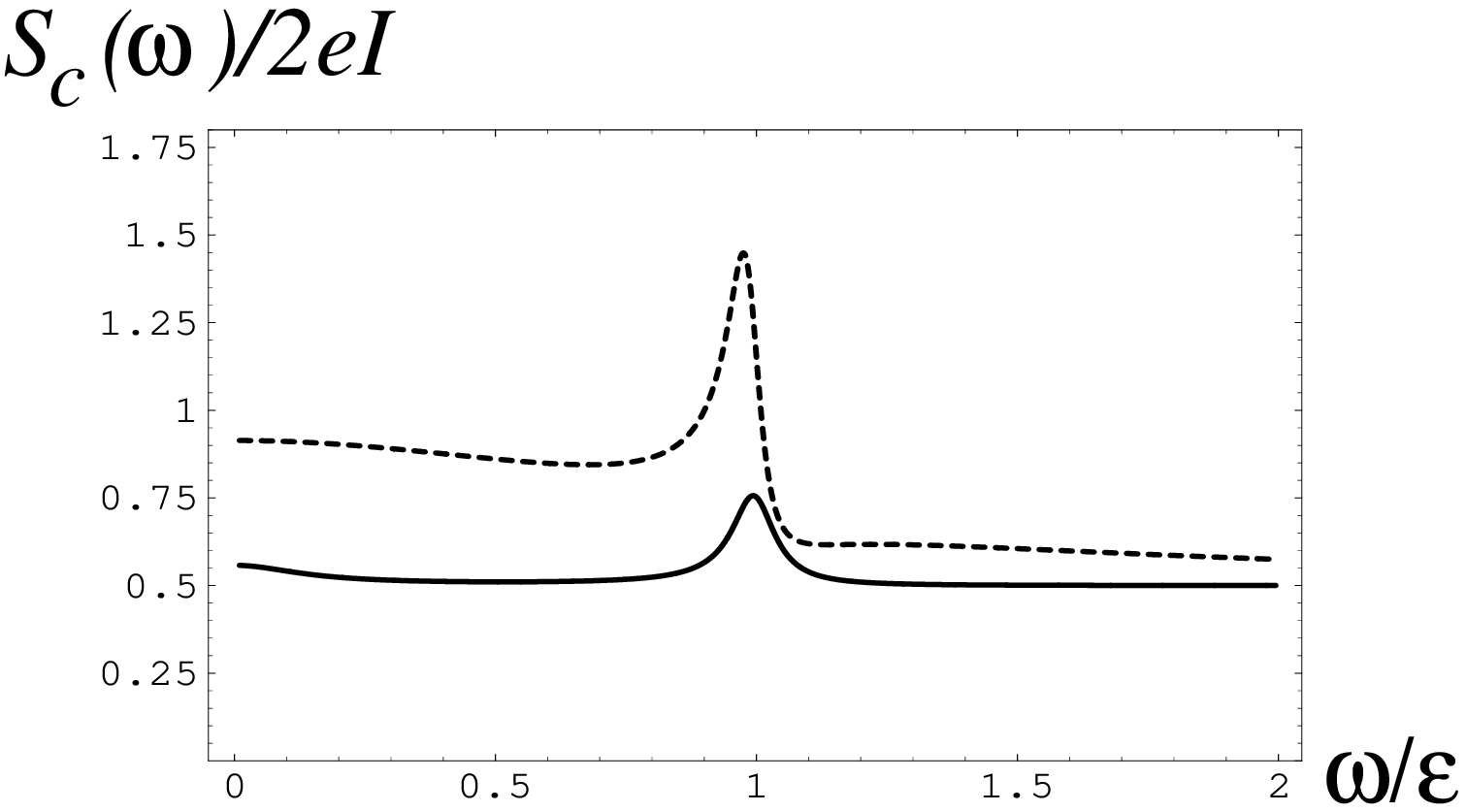}
\end{center}
{\begin{small} Fig.~7. The Fano factor for the circuit ($\alpha
=\beta =1/2$) versus $\omega $ for a polarized electron current
with ferromagnetic reservoirs and Coulomb blockade and
$\theta_L,\theta_R=\pi/2$. The solid line corresponds to
$\Gamma_L=\Gamma_R=0.1\epsilon$ and the dashed line to
$\Gamma_L=\epsilon$ and $\Gamma_R=0.1\epsilon$.
\end{small}}
\end{minipage} \\ \\

One finds from Figs.~6 and 7 that the Coulomb blockade modifies
the current spectrum very drastically with respect to the
non-interacting case, Fig.~5.

\section{Conclusions}

In this paper, we study the interference effects in
magneto-transport through Zeeman split levels of quantum dots or
impurities. We concentrated on the time-dependent properties and
the power spectrum of the electric current by applying the new
approach using quantum rate equations, which is mostly suitable
for this type of problems. We explicitly demonstrated that our
method produces the same results as a single electron approach,
widely used for a description of non-interacting electron
transport. Yet the quantum rate equations method is valid also for
the case of interacting electrons and accounts for the Coulomb
blockade in the most simple and precise way.

Our results indicate that the Coulomb blockade plays an important
role in the spectral properties of the transport current. First of
all, in the presence of Coulomb blockade the signal-to-noise ratio
significantly is amplified, as one can observe from the results in
the previous sections. This is probably a consequence of the
prohibition of double occupation of the resonant level in the
quantum dot. Indeed, when two electrons in the dot are present,
the interference effects are suppressed due to the
``randomization'' of the relative phase. Interestingly, the dip in
the noise spectrum for the noninteracting  electrons is replaced
by a peak as result of Coulomb interaction. Clearly the Coulomb
interaction modifies the phase of the electrons tunneling trough
the dot, which ``flips'' the spectral feature in the noise. The
details of this very interesting phenomena must be studied in the
future.

We emphasize that the coherent oscillations in the current can be
observed only for polarized current and that oscillations
disappear for unpolarized current. This is different from the
resonant transport through two orbital levels of a quantum dot or
impurity, where the quantum interference effects can be observed
even in unpolarized case. Therefore it is most natural to use
ferromagnetic leads for observation and utilization of quantum
interference effect in the magneto-transport. Thus our
calculations were mostly concentrated on this case. Our results
show explicitly the appearance of peak or dip at a frequency near
the Zeemann splitting frequency (Larmour frequency). We believe
that this phenomenon can be useful for analyzing the noise
spectroscopy of quantum dots or impurities. Indeed, the Zeeman
splitting of a localized quantum dot orbital must be sensitive to
local magnetic fields, and therefore one can hope that such
coherent effect, if observed experimentally, may allow for
detection of the local hyperfine structure of the dot/impurity.
This, however, must be a subject of a separate investigation.
\section{Acknowledgement}
We thank J. Brown, L. Fedichkin, M. B. Hastings, M. Hawley, and I.
Martin for valuable discussions. We are especially grateful to
Gary D. Doolen for important remarks and for proofreading the
manuscript. The work was supported by the Department of Energy
under Contract No. W-7405-ENG-36 and by DOE Office of basic Energy
Sciences. D. M. was supported, in part, by the US NSF grant
DMR-0121146.

\appendix
\section{Single-electron description}
In the case of non-interacting electrons one can compare our
results with those obtained using a single-electron approach.
Although the latter is widely used in the literature, it is
usually restricted to the time-independent (stationary) case. Here
we present an extension of the single-electron approach for the
non-stationary case. This would allow us to evaluate the
time-dependent resonant current, Fig.~3, and to compare the
results with those obtained from Eqs.~(\ref{b3}).

Let us consider a system consisting of the reservoirs and the
quantum dot filled with only a single electron. We assume that
this electron is initially in the left reservoir (emitter) at the
level $E_{\bar l}$ with the spin polarized along the {\bf
n}-direction, Fig.~1. The electron motion is described by a wave
function which can be written in the most general way as\cite{g1}:
\begin{equation}
  |\Psi (t)\rangle =\left [\sum_{l,s} \sb_{ls}(t)a_{ls}^\dagger
  + \sum_{d=1,2} \sb_{d}(t)a_{d}^\dagger +
\sum_{r,s'} \sb_{rs'}(t)a_{rs'}^\dagger \right ]|0\rangle\, ,
\label{a2}
\end{equation}
where $\sb_\alpha (t)$ is the amplitude of finding the electron in
the state $\alpha$ given by a corresponding creation operator.
These amplitudes are obtained from the Schr\"odinger equation
$|\Psi (t)\rangle$, with the initial conditions
$\sb_{ls}(0)=\delta_{l,\bar l}\delta_{s,1/2}$ and
$\sb_d(0)=\sb_{rs}(0)=0$. It is useful to use the Laplace
transform, $\tilde \sb(E)=\int_0^\infty \sb(t)\exp (iEt)dt$. In
this case the time-dependent Schr\"odinger equation for the
amplitudes $\tilde \sb(E)$ becomes the following system of linear
algebraic equations
\begin{mathletters}
\label{a4}
\begin{eqnarray}
&&(E-E_{ls})\tilde \sb_{ls}(E)
-\Omega_{l}\sum_{d'=1,2}\sd^{(1/2)}_{s,s_{d'} }(\theta_L)\tilde
\sb_{d'}(E) =i\delta_{l,\bar l}\delta_{s,1/2}
\label{a4a}\\
&&(E-E_{d})\tilde
\sb_{d}(E)-\sum_{l,s}\Omega_{l}\sd^{(1/2)}_{s_d,s}(\theta_L) \tilde
\sb_{ls}(E)
-\sum_{r,s'}\Omega_{r}\sd^{(1/2)}_{s_d,s'}(\theta_R)\tilde
\sb_{rs'}(E)=0
\label{a4b}\\
&&(E-E_{rs'})\tilde \sb_{rs'}(E)
-\Omega_{r}\sum_{d'=1,2}\sd^{(1/2)}_{s',s_{d'}}(\theta_R)\tilde
\sb_{d'}(E)=0\, . \label{a4c}
\end{eqnarray}
\end{mathletters}
Substituting $\tilde \sb_{ls}$ and $\tilde \sb_{rs'}$ from
Eqs.~(\ref{a4a}), (\ref{a4c}) into Eq.~(\ref{a4b}) and replacing the
sums on $l$ and $r$ by the integrals, we obtain
\begin{mathletters}
\label{a5}
\begin{eqnarray}
&&\left (E-E_{1}+i{\Gamma_L+\Gamma_R\over 2}\right ) \tilde
\sb_{1}(E)=i{\Omega_L\cos (\theta_L/2)\over E-E_{\bar l,1/2}}
\label{a5a}\\
&&\left (E-E_{2}+i{\Gamma_L+\Gamma_R\over 2}\right ) \tilde
\sb_{2}(E)=-i{\Omega_L\sin (\theta_L/2)\over E-E_{\bar l,1/2}}
\label{a5b}\, .
\end{eqnarray}
\end{mathletters}
Note that the amplitudes, $\tilde \sb_{1}(E)$ and $\tilde
\sb_{2}(E)$, are decoupled in Eqs.~(\ref{a5}) although the
corresponding states are connected via the continuum. The reason
is that the spin-flip couplings of the dot with the reservoirs are
of the opposite sign for the spin-up and the spin-down states of
the dot ($E_1$ and $E_2$ in Fig.~1). However, for the general case
of resonant tunneling through two levels, the corresponding
amplitudes are coupled via the interaction through
continuum\cite{ieee}.

Using the inverse Laplace transform $\sb_{1,2}(t)=\int\tilde
\sb_{1,2}(E) \exp (-iEt)dE/(2\pi)$, we obtain for the amplitudes,
$\sb_{1,2}(t)$, for finding the electron inside the dot
\begin{mathletters}
\label{a9}
\begin{eqnarray}
\sb_{1}(t)&=&{\Omega_L\cos (\theta_L/2)\over E_L-E_{1}+i{\Gamma\over
2}} \left (e^{-iE_Lt}-e^{-iE_{1}t-{\Gamma\over 2}t}\right )
\label{a9a}\\
\sb_{2}(t)&=&-{\Omega_L\sin (\theta_L/2)
  \over E_L-E_{2}+i{\Gamma\over 2}}
\left (e^{-iE_Lt}-e^{-iE_{2}t-{\Gamma\over 2}t}\right )\, ,
\end{eqnarray}
\end{mathletters}
where $\Gamma=\Gamma_L+\Gamma_R$. The probability amplitude of
finding the electron inside the collector is $\tilde
\sb_{rs'}(t)=\int\tilde \sb_{rs'}(E) \exp (-iEt)dE/(2\pi)$, where
$\tilde \sb_{r,s'}(E)$ is given by Eq.~(\ref{a4c})
\begin{equation}
  \tilde \sb_{rs'}(E)={\Omega_R\over E-E_{rs'}}\sum_{d}\sd^{(1/2)}_{s's_d}(\theta_R)
\tilde \sb_{d}(E)\, . \label{a7}
\end{equation}

The above equations determine the motion of a single electron
placed initially in the emitter. In order to obtain the polarized
current, $I_{s'}(t)$, in the single-electron model one has to sum
over all initially occupied states $E_{\bar l}$ of the emitter and
over all available states $E_r$ of the collector. Thus $I_{s'}
=dN_{s'}(t)/dt$, where $N_{s'}(t)=\sum_{\bar l,r}|\sb_{rs'}(t)|^2$
is the average number of electrons with spin-up and spin-down
($s'=\pm 1/2$), accumulated in the collector by the time $t$.
 Using the inverse Laplace transform and
replacing $\sum_{\bar l,r}\to\int\rho_L\rho_RdE_LdE_R$ we obtain
\begin{equation}
  N_{s'}(t)=\int \rho_L\rho_RdE_LdE_R\int{dEdE'\over (2\pi )^2}\tilde
  \sb_{rs'}(E)\tilde \sb^*_{rs'}(E')e^{i(E'-E)t}
\label{a6}
\end{equation}

Substituting Eq.~(\ref{a7}) into Eq.~(\ref{a6}) and integrating over
$E_{rs'}$ one obtains for the polarized current
\begin{mathletters}
\label{a8}
\begin{eqnarray}
  I_{1/2}(t)&=&\Gamma_R\int_{\mu_R}^{\mu_L} \rho_LdE_L|\cos
(\theta_R/2)\sb_{1}(t) -\sin (\theta_R/2)\sb_{2}(t)|^2
\\[5pt]
\label{a8a}
 I_{-1/2}(t)&=&\Gamma_R\int_{\mu_R}^{\mu_L} \rho_LdE_L|\sin
(\theta_R/2)\sb_{1}(t) +\cos (\theta_R/2)\sb_{2}(t)|^2 \label{a8b}
\end{eqnarray}
\end{mathletters}
where the amplitudes $\sb_{1,2}(t)$ are given by Eqs.~(\ref{a9}).
Note that these amplitudes in the stationary limit,
$\sb_{1,2}(t\to\infty )$, are the transmission amplitudes
describing the resonance tunneling through the levels $E_{1,2}$
\cite{ped}. Thus Eqs.~(\ref{a8}) represent a generalization of the
Landauer formula for the time-dependent case.

For large bias, $\mu_L-\mu_R\gg \Gamma$, the integration over
$E_L$ in Eqs.~(\ref{a8}) can be performed analytically using
Eqs.~(\ref{a9}) for the amplitudes $\sb_{1,2}(t)$. As a result we
finally arrive at Eq.~(\ref{a13}) obtained from Eq.~~(\ref{b3})
for the case of non-interacting electrons. This agreement with the
case of non-interacting electrons is quite remarkable since our
rate equations dealing with many-electron states are very
different from those obtained in the single electron framework.
Yet, this is not surprising since in the case of non-interacting
electrons the single electron description is valid. In fact,
Eqs.~(\ref{a5}) can be mapped to Eq.~(\ref{b6}) using
$|\sb_i(t)|^2=\sigma_{ii}(t)+\sigma_{33}(t)$, where $i=1,2$ and
$\sb_1(t)\sb^*_2(t)=\sigma_{12}(t)$.

\end{document}